\documentclass[prl,aps,nobalancelastpage,showpacs,amssymb,twocolumn,citeautoscript,preprintnumbers,floatfix]{revtex4}

\usepackage{mathrsfs}
\usepackage{amsmath}
\usepackage{amssymb}
\usepackage{graphicx}% Include figure files
\usepackage{color}
\usepackage{dcolumn}% Align table columns on decimal point
\usepackage{bm}% bold math

\makeatletter

\newcommand{\Rmnum}[1]{\expandafter\@slowromancap\romannumeral #1@}

\makeatother

\begin{document}

\title{Pressure-enhanced superconductivity in Eu$_3$Bi$_2$S$_4$F$_4$}

\author{Yongkang Luo$^{1}$\footnote[1]{Electronic address: ykluo@lanl.gov}, Hui-Fei Zhai$^{2}$, Pan Zhang$^{2}$, Zhu-An Xu$^{2}$, Guang-Han Cao$^{2}$\footnote[2]{Electronic address: ghcao@zju.edu.cn}, and J. D. Thompson$^{1}$}
\address{$^1$Los Alamos National Laboratory, Los Alamos, New Mexico 87545, USA;}
\address{$^2$Department of Physics, Zhejiang University, Hangzhou 310027, China.}

\date{\today}

\begin{abstract}

The pressure effect on the newly discovered charge-transferred BiS$_2$-based superconductor, Eu$_3$Bi$_2$S$_4$F$_4$, with a $T_c$ of 1.5 K at ambient pressure, is investigated by transport and magnetic measurements. Accompanied with the enhancement of metallicity under pressures, the onset superconducting transition temperature increases abruptly around 1.0 GPa, reaching $\sim$10.0 K at 2.26 GPa. AC magnetic susceptibility measurements indicate that a new superconducting phase with a higher $T_c$ emerges and dominates at high pressures. In the broad pressure window of 0.68 GPa$\leq$$p$$\leq$2.00 GPa, the high-$T_c$ phase coexists with the low-$T_c$ phase. Hall effect measurements reveal a significant difference in electronic structures between the two superconducting phases. Our work devotes the effort to establish the commonality of pressure effect on the BiS$_2$-based superconductors, and also uncovers the importance of electron carrier density in the high-$T_c$ phase.

\end{abstract}

%74.62.Fj Effects of pressure
%61.50.Ks Crystallographic aspects of phase transformations; pressure effects
%71.28.+d Narrow-band systems; intermediate-valence solids
%74.70.Dd Ternary, quaternary, and multinary compounds

\maketitle

%\section{Introduction}

The layered bismuth-sulfide superconductors have recently attracted much attention after the generations of cuprates\cite{Bednorz-La2CuO4_Sr} and iron-pnictides\cite{Hosono-La1111_F}. Following the discovery of superconductivity (SC) in Bi$_4$O$_4$S$_3$ ($T_c$=8.6 K)\cite{Mizuguchi-Bi4O4S3}, SC was also discovered in $Ln$BiS$_2$O$_{1-x}$F$_{x}$ ($Ln$=La, Ce, Pr, Nd, Yb)\cite{Mizuguchi-La1121_F,WenHH-Ce1121_F,Jha-Pr1121_F,Yazici_Ln1121_F,Demura-Nd1121_F}, LaBiSe$_2$O$_{1-x}$F$_{x}$\cite{Maziopa-LaSe1121_F,WenHH-LaSe1121pre}, and the structural analogs Sr$_{1-x}$$Ln_{x}$BiS$_2$F ($Ln$=La, Ce, Pr, Nd, Sm)\cite{LiYK-Sr1121_La,Jha-Sr1121_Ln} and EuBiS$_2$F\cite{Zhai-Eu1121}. Besides chemical doping, pressure effect is another promising way to search for and/or enhance SC in BiS$_2$-based compounds. Taking LaBiS$_2$O$_{0.5}$F$_{0.5}$ as an example, the sample grown at atmospheric pressure becomes superconducting below $\sim$3.3 K. While under pressure, $T_c$ initially increases via a transient process, reaches a maximum value of $\sim$10.1 K at around 1.2 GPa, and then gradually decreases with increasing pressure\cite{Maple-LaCe1121pre}. %Similar result was reported for the high-pressure annealed LaBiS$_2$O$_{0.5}$F$_{0.5}$ sample\cite{Kotegawa-BiS2pre}.
High pressure X-ray diffraction (XRD) measurements suggested a structural phase transition from tetragonal ($P4/nmm$) to monoclinic ($P2_1/m$) at $\sim$0.8 GPa, which is probably associated with the pressure-induced enhancement of SC\cite{Tomita-La1121pre}.

Recently, Zhai {\it et al.} reported the anomalous Eu-valence state and superconductivity in ``undoped" Eu$_3$Bi$_2$S$_4$F$_4$\cite{Zhai-Eu3244}. This compound crystallizes in a tetragonal structure with $I4/mmm$ (No. 139) space group. The Eu$_3$F$_4$-layers and BiS$_2$-bilayers stack alternately along $\textbf{c}$-axis, developing a highly two-dimensional (2D) crystalline structure.  There are three Eu ions in one unit cell occupying two inequivalent crystallographical sites: two of the Eu ions [denoted by Eu(1)] are essentially divalent and form an antiferromagnetic (AFM) ordering below $T_N$=2.3 K; while the third Eu ion [Eu(2)] has an average valence of $\sim$$+$2.6. This intermediate Eu-valence state donates electrons into the conducting BiS$_2$-layers. SC with $T_c$=1.5 K was achieved in this self-doped material. One intuitive attempt is to tune the superconductivity as well as Eu valence by means of pressure effect. Since an external pressure potentially strengthens the hybridization between Eu-$4f$ and Bi-$6p$ orbitals, a higher $T_c$ is possible if more charges are transferred into the BiS$_2$-layers.

In this paper, we report the hydrostatic pressure effect on Eu$_3$Bi$_2$S$_4$F$_4$ by electrical resistivity, Hall effect and AC magnetic susceptibility measurements. We find that the Eu valence is essentially unchanged under pressures up to 2.26 GPa, and that the $T$ vs. $p$ electronic phase diagram basically resembles that of $Ln$BiS$_2$O$_{0.5}$F$_{0.5}$\cite{Maple-LaCe1121pre,Maple-Ln1121pre}. The onset of superconducting transition jumps from $\sim$1.5 K to $\sim$10.0 K at the pressure around 1.00 GPa, in the vicinity of which a new high-$T_c$ SC phase emerges. Both of the two SC phases are of bulk nature, and they coexist in a wide pressure range. The Eu(1) AFM ordering becomes indistinguishable near the onset of this new SC phase. A significant difference in electronic structures between these two SC phases is revealed by Hall effect measurement.

%\section{Experimental details}

A poly-crystalline Eu$_3$Bi$_2$S$_4$F$_4$ sample of high quality was synthesized by solid state reaction as described elsewhere\cite{Zhai-Eu3244}. By controlling a suitable S deficiency in the nominal composition ($\sim$2.5\%), we were able to reduce the Bi$_2$S$_3$ and EuF$_{2.4}$ impurity phases, and thus further improved the sample purity. A piston-clamp type pressure cell was used to pressurize the sample, and Daphne oil 7373 was employed as a pressure-transmitting medium. Hydrostatic pressure up to 2.26 GPa was applied in the experiment, during which highly pure Pb was used as the manometer. Ohmic contacts were made in a Hall-bar geometry, and both (magneto-)resistivity and Hall resistivity were measured by an LR-700 AC resistance bridge. AC magnetic susceptibility measurements were performed by a set of handmade coils, which consists of a drive coil, a pick-up coil and a compensation coil. A drive current of 1 mA at 137.37 Hz was applied in the measurement, and the signal was detected by an SR-850 lock-in amplifier.

%\section{Results and Discussion}

\begin{figure}[htbp]
\includegraphics[width=8.0cm]{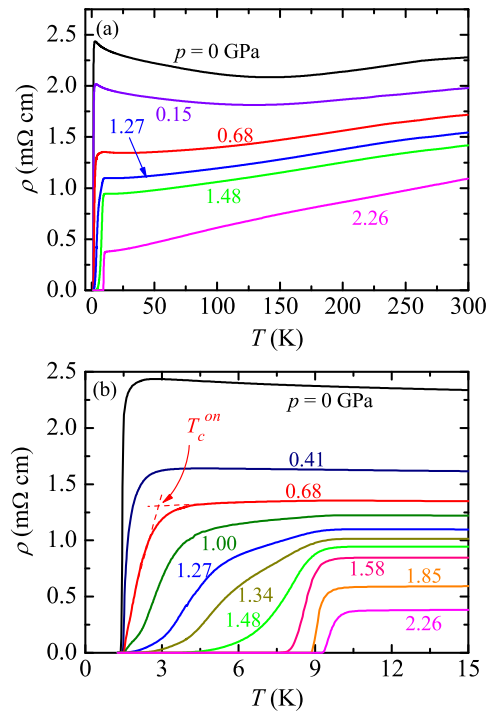}
\caption{(Color online)\label{Fig.1} Temperature dependent resistivity of Eu$_3$Bi$_2$S$_4$F$_4$ at various pressures. Panel (a) shows data over the full temperature range; whereas, panel (b) displays an enlarged plot of $\rho(T)$ below 15 K. Also shown in (b) is the definition of the onset temperature $T_c^{on}$.}
\end{figure}

Fig.~\ref{Fig.1}(a) displays several representative $\rho(T)$ curves measured at various pressures. At ambient pressure, $\rho(T)$ shows a metallic-like temperature dependence above 140 K, below which $\rho(T)$ turns up that was attributed to Anderson localization (probably due to S deficiency)\cite{Zhai-Eu3244}. We emphasize that the sample investigated in this work is better in phase purity than the one reported in our previous paper\cite{Zhai-Eu3244}, confirmed by XRD (data not shown). Although the upturn in $\rho(T)$ at low temperature looks more pronounced, the magnitude of resistivity for the whole temperature range is lower. The material shows a sharp superconducting transition at $T_c$=1.5 K at ambient pressure, in good agreement with the previous report. In general, metallicity of Eu$_3$Bi$_2$S$_4$F$_4$ is greatly enhanced under pressure, and meanwhile the semiconducting behavior is gradually suppressed. In Fig.~\ref{Fig.1}(b) we present an expanded view of the low temperature part to show the superconducting transition more clearly. For $p$$<$1.58 GPa, the superconducting transition width broadens significantly, whilst for higher pressures, the transition sharpens again at about 9 K. Herein we take the onset of SC transition [$T_c^{on}$ illustrated in Fig.~\ref{Fig.1}(b)] as the characteristic temperature, which has been widely adopted in most of the literature on BiS$_2$-based superconductors. For the highest pressure we have achieved, $p$=2.26 GPa, $T_c^{on}$ reaches 10.0 K, which is comparable to the highest record ($T_c^{on}$=10.7 K for LaBiS$_2$O$_{0.5}$F$_{0.5}$ under an optimized pressure\cite{Tomita-La1121pre}) in BiS$_2$-based superconductors. Obviously, $T_c^{on}$ is still going up with pressure. This implies that the $T_c^{on}$ value could be further elevated at higher pressures.

\begin{figure}[htbp]
\includegraphics[width=8.5cm]{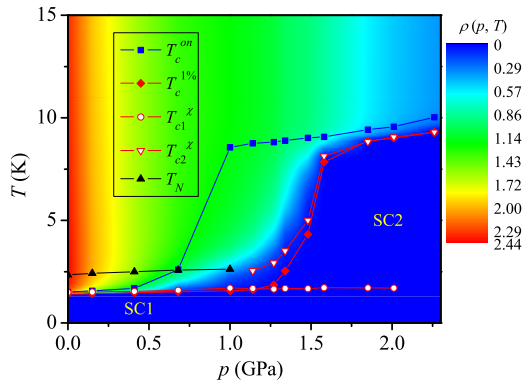}
\caption{(Color online)\label{Fig.2} Phase diagram of Eu$_3$Bi$_2$S$_4$F$_4$, contour-plotted with $\rho(p, T)$. The blue region on the bottom marks superconductivity. $T_c^{on}$ characterizes the onset of SC transition in $\rho(T)$; $T_c^{1\%}$ represents the temperature below which the resistivity is smaller than 1\% of that in the normal state; $T_{c1}^{\chi}$ and $T_{c2}^{\chi}$ stand for the transitions observed in $\chi'(T)$ corresponding to SC1 and SC2 phases respectively, seeing Fig.~\ref{Fig.3}; while $T_N$ is the temperature scale of the long-range AFM ordering of Eu(1) moments.}
\end{figure}

We summarize the resistivity results by presenting a contour plot of $\rho(p, T)$ shown in Fig.~\ref{Fig.2}. The blue region on the bottom of Fig.~\ref{Fig.2} characterizes SC. One clearly finds the broadened SC transition under moderate pressures by looking at the light-blue part on the border of the SC region. It is worthwhile to point out that the pressure effect reported here resembles that observed in most $Ln$BiS$_2$O$_{0.5}$F$_{0.5}$ ($Ln$=La,Ce,Pr,Nd) systems\cite{Maple-LaCe1121pre,Maple-Ln1121pre,WenHH-LaSe1121pre,Tomita-La1121pre}. We may remark the common feature that a new SC phase with a higher $T_c$ emerges at  high pressure for BiS$_2$-based superconductors. For clarity, hereafter the low-$T_c$ SC phase is called {\it SC1}, while the high-$T_c$ SC phase is named as {\it SC2}. In our result, both SC1 and SC2 are enhanced under pressure; this is in contrast to the $Ln$BiS$_2$O$_{0.5}$F$_{0.5}$ systems\cite{Maple-LaCe1121pre,Maple-Ln1121pre,WenHH-LaSe1121pre}, in which $T_c$ immediately starts to decrease after SC2 is stabilized at high pressure. This discrepancy is probably because the self-doped Eu$_3$Bi$_2$S$_4$F$_4$ resides in the under-doped regime [$\sim$30\%, according to the Eu(2) valence $\sim$$+$2.6\cite{Zhai-Eu3244}], while the 50\% Fluorine-doped $Ln$BiS$_2$O is optimally doped. It should be pointed out that the relationship between SC1 and SC2 seems like phase separation. The SC2 phase appears at the pressure $\sim$1.00 GPa [see in Fig.~\ref{Fig.1}(b)], with $T_c^{on}$=8.56 K, while the zero-resistivity temperature (defined by $T_c^{1\%}$=1.57 K) at this pressure is still low. This reminds us that the volume fraction of SC2 phase is very small on its birth-point. As pressure increases, the SC2 phase grows, and superconducting current path becomes well established, which sharpens the SC transition. More evidences will be provided in AC magnetic susceptibility measurements.

\begin{figure}[htbp]
\includegraphics[width=8.0cm]{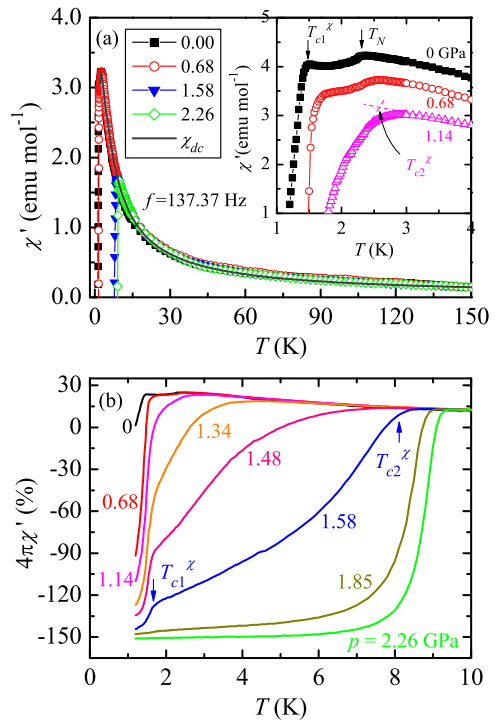}
\caption{(Color online)\label{Fig.3} The real part of AC magnetic susceptibility $\chi'(T)$ under pressures, measured at a frequency $f$=137.37 Hz. $\chi_{dc}$ is the DC magnetic susceptibility measured with SQUID at atmosphere pressure. The inset is a zoom-in plot of the low temperature region, the data of which have been vertically shifted for clarity. (b) Diamagnetic volume fraction (4$\pi\chi'$) as a function of $T$. Note that the demagnetizaion factor is not taken into account. }
\end{figure}

Turning now to the AC magnetic susceptibility. The real part of the voltage signal, after being subtracted by the background (measured from the empty coil), was calibrated by the DC magnetic susceptibility $\chi_{dc}$ ($p$=0, see Ref.\cite{Zhai-Eu3244}). The real part of the AC susceptibility $\chi'$ are shown in Fig.~\ref{Fig.3} as a function of $T$. In the normal state, within the data resolution, $\chi'(T)$ for all the pressures almost collapse onto a single Curie-Weiss-like curve, strongly demonstrating that the valence state of Eu(2) essentially remains unchanged under pressures up to 2.26 GPa. In the inset to Fig.~\ref{Fig.3}(a), we show the low temperature part of $\chi'(T)$. For $p$=0, the peak at $T_N$=2.35 K corresponds to the AFM transition of Eu(1) moments. Below $T_{c1}^{\chi}$=1.49 K, $\chi'(T)$ starts to drop rapidly, indicating the bulk SC1 phase transition. With increasing pressure, both $T_N$ and $T_{c1}^{\chi}$ increase slightly, and meanwhile the diamagnetic Meissner effect below $T_{c1}^{\chi}$ becomes more pronounced. At $p$=1.14 GPa, $\chi'(T)$ starts to deviate from the Curie-Weiss-like behavior around $T_{c2}^{\chi}$=2.54 K, while no clear AFM transition can be distinguished. Combined with the resistivity, $T_{c2}^{\chi}$ should be ascribed to the SC2 phase transition. Apparently, both SC1 and SC2 phases are bulk properties, and the two phases coexist in a wide pressure range. $T_{c2}^{\chi}$ increases quickly in the moderate pressure region 1.14 GPa$\leq$$p$$\leq$1.58 GPa, and finally slows down at the high pressure region, seeing Fig.~\ref{Fig.3}(b). At $p$=2.26 GPa, a complete Meissner effect is observed below $T_{c2}^{\chi}$, while no clear sign of SC1 phase can be seen. The diamagnetic volume fraction 4$\pi\chi'$ at this pressure exceeds 150\% because we did not take into account the demagnetization factor.

We now fulfill the phase diagram of the pressure effect on Eu$_3$Bi$_2$S$_4$F$_4$, as shown in Fig.~\ref{Fig.2}. It is clearly seen that $T_c^{1\%}(p)$ overlaps with $T_{c1}^{\chi}(p)$ at pressures up to 1.14 GPa. After a crossover-like increase between 1.14 GPa and 1.58 GPa, $T_{c}^{1\%}(p)$ coincides with $T_{c2}^{\chi}(p)$ in the high pressure region. Since $T_{c1}^{\chi}$ and $T_{c2}^{\chi}$ are the characteristic temperature scales for bulk SC1 and SC2 transitions, this pressure dependent $T_c^{1\%}$ supports the phase separation picture. We also notice a salient feature on this phase diagram, that the long-range Eu(1) AFM transition becomes indistinguishable near the pressure where the SC2 phase emerges. Intuitively, the AFM ordering is not favored in the SC2 phase, probably because there is a competition between the SC condensation and the Ruderman-Kittel-Kasuya-Yosida (RKKY) interaction that mediates the Eu(1) AFM ordering. On the onset of SC2 phase, it forms like a small droplet. {\it Microscopically}, these SC2 droplets randomly distribute inside the bulk sample, which is likely to break the long range Eu(1) AFM ordering. However, based on the present measurements, we can not exclude the possibility that the Eu(1) AFM ordering, if it still survives in the SC1 phase under pressure, is masked by the predominant SC signal from the SC2 phase. More sensitive and microscopic techniques e.g. $\mu$SR and neutron scattering experiments are required to solve this problem.

\begin{figure}[htbp]
\includegraphics[width=8.0cm]{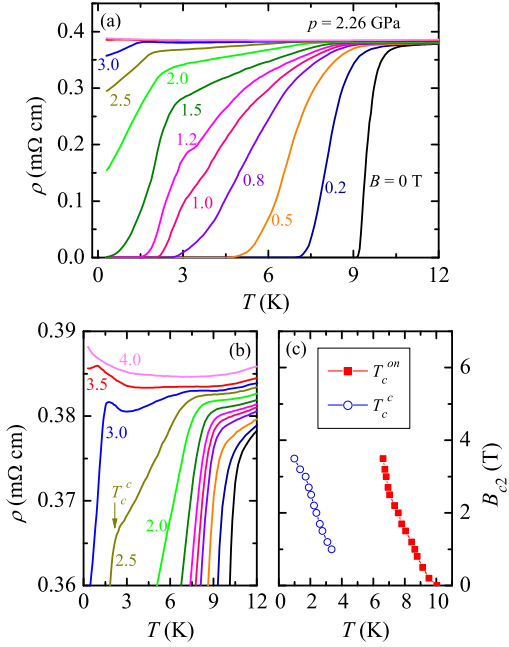}
\caption{(Color online)\label{Fig.4} Panel (a), superconducting transition under various magnetic fields at the pressure $p$=2.26 GPa. Panel (b) is an enlarged plot of panel (a). Panel (c) shows the superconducting $B_{c2}$-$T$ phase diagram at $p$=2.26 GPa. }
\end{figure}

At the highest pressure $p$=2.26 GPa in our study, where it is believed to be a nearly pure phase of SC2, we systematically studied the electrical resistivity under various magnetic fields. The data are shown in Fig.~\ref{Fig.4}(a). All these measurements were made after a field-cooling process. With increasing field, $T_c^{1\%}$ is drastically suppressed and the zero-resistivity state can not be observed above 1.5 T for $T$$\geq$0.3 K. Whereas $T_c^{on}$ seems more robust against an external magnetic field, as seen in Fig.~\ref{Fig.4}(b). When $B$ exceeds 1.0 T, we observe an additional anomaly in $\rho(T)$ curve near 3 K. A similar phenomenon was also observed in LaBiS$_2$O$_{0.5}$F$_{0.5}$, and was interpreted as the anisotropic upper critical field ($B_{c2}$)\cite{Mizuguchi-La1121Hc2}. The high anisotropy of $B_{c2}$ in LaBiS$_2$O$_{0.5}$F$_{0.5}$ was also confirmed by field rotation experiments on single crystalline samples\cite{Nagao-Ln1121SX}, which revealed that $B_{c2}$ in the $\textbf{B}$$\parallel$$\textbf{c}$ direction is much smaller than that for $\textbf{B}$$\parallel$$\textbf{ab}$ direction. Due to the more 2D crystallographic structure of Eu$_3$Bi$_2$S$_4$F$_4$, it is likely that this anisotropy of $B_{c2}$ in Eu$_3$Bi$_2$S$_4$F$_4$ would be even higher than that in LaBiS$_2$O$_{0.5}$F$_{0.5}$. Therefore, it is reasonable that the superconducting current path in those grains with the orientation of $\textbf{c}$$\parallel$$\textbf{B}$ can be destroyed at a temperature [defined as $T_c^c$ in Fig.~\ref{Fig.4}(b)] that is lower than $T_c^{on}$. In Fig.~\ref{Fig.4}(c), we summarize the $B_{c2}$ vs. $T$ phase diagram. Concave dependencies are clearly seen, in both $T_c^{on}$ and $T_c^c$ criterions, which is a signature of multi-band effect. The $B_{c2}(0)$ for $\textbf{B}$$\parallel$$\textbf{c}$ is estimated to be $\sim$4.0 T; while for $\textbf{B}$$\parallel$$\textbf{ab}$, $B_{c2}(0)$ should be much higher but is hard to estimate.

\begin{figure}[htbp]
\includegraphics[width=8.0cm]{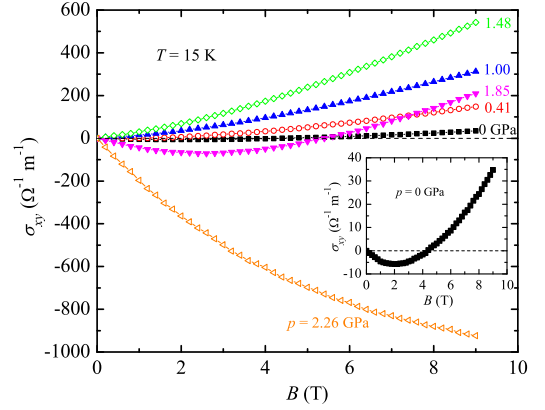}
\caption{(Color online)\label{Fig.5} Field dependent Hall conductivity $\sigma_{xy}(B)$ under various pressures, measured at 15 K. The data for $p$=0 are also shown in the inset for clarity. }
\end{figure}

Finally, we also measured the Hall effect under pressure. The measurement was carried out at 15 K, well above the maximum $T_c^{on}$. Since the metallicity of the sample is greatly enhanced under pressure, for comparison, we have converted Hall resistivity ($\rho_{yx}$) to Hall conductivity ($\sigma_{xy}$) via $\sigma_{xy}$=$\frac{\rho_{yx}}{\rho^2+\rho_{yx}^2}$. The field dependent $\sigma_{xy}$ for various pressures are displayed in Fig.~\ref{Fig.5}. For all the pressures, $\sigma_{xy}(B)$ is severely non-linear, implying the multi-band effect. At ambient pressure, $\sigma_{xy}(B)$ is negative under low fields, but undergoes a sign change at $B$=4.2 T, seeing the inset to Fig.~\ref{Fig.5}. Such a profile of $\sigma_{xy}(B)$ indicates the coexistence of electron-type minority carriers with high mobility and hole-type majority carriers with low mobility. In fact, earlier first-principles calculations on the analog EuBiS$_2$F [based on a local spin density approximation with on-site Coulomb interaction (LSDA+$U$)] manifested that there are three bands across the Fermi level: one hole band near the $M$-point and two electron bands around the $X$-point\cite{Zhai-Eu1121}. The hole band is from Eu(2)-$4f$ orbitals, and thus is believed to be of low mobility. This hole band donates electrons to the electron bands (mainly of Bi-$6p$ characteristic\cite{Zhai-Eu1121,Usui-band}), leaving the conduction bands in EuBiS$_2$F partially filled with transferred charges. The situation could be similar in Eu$_3$Bi$_2$S$_4$F$_4$, if we consider that the Eu$(1)$F$_2$-layer only serves as an ``insulating" block\cite{Zhai-Eu3244}. As pressure increases, $\sigma_{xy}(B)$ increases and becomes completely positive in the mixture of SC1 and SC2. For $p$$>$1.48 GPa where the SC2 phase is well developed, $\sigma_{xy}(B)$ drops again, and becomes electron-dominant at $p$=2.26 GPa. Such an evolution of $\sigma_{xy}(B)$ with pressure is hard to interpret quantitatively. However, if we compare the $\sigma_{xy}(B)$ curves at $p$=0 and 2.26 GPa, which can be regarded to be of pure SC1 and SC2 phases respectively, we may remark that a drastic change in band structure and Fermi surface topology takes place when it changes from SC1 phase to SC2 phase. Recently, Tomita {\it et al.} reported a structural phase transition in LaBiS$_2$O$_{0.5}$F$_{0.5}$ between the ambient-pressure and high-pressure states\cite{Tomita-La1121pre}. Further density functional theory (DFT) calculations confirmed that a structural phase transition from a tetragonal $P4/nmm$ phase at ambient pressure to a monoclinic $P2_1/m$ phase at high pressure occurs upon pressurizing. A similar structural phase transition could also happen in Eu$_3$Bi$_2$S$_4$F$_4$ under pressure, and a new superconducting phase (namely the SC2 phase) with a higher $T_c$ is favored in this modified crystalline structure. And moreover, because the RKKY interaction depends on electrons near as well as deep inside the Fermi sea, suppression of the RKKY interaction by SC allows additional electrons to participate in SC condensation. This may result in a higher $T_c$ in the SC2 phase of Eu$_{3}$Bi$_2$S$_4$F$_4$, and other BiS$_2$-based superconductors as well. Of course, to clarify this issue, more theoretical and experimental researches, especially single crystal synthesis, are deserved.

%\section{Conclusion}

In conclusion, we have pressurized and investigated the transport and magnetic properties of the charge transferred superconductor Eu$_3$Bi$_2$S$_4$F$_4$. With $T_c^{on}$ abruptly increasing from $\sim$1.5 K to $\sim$10.0 K in the pressure window 0.68 GPa$\leq$$p$$\leq$1.00 GPa, and a crossover of $T_c^{1\%}(p)$ from $T_{c1}^{\chi}$ to $T_{c2}^{\chi}$, the results demonstrate that a new superconducting phase emerges at high pressures. The two SC phases are phase-separated and coexist in the bulk sample. Hall effect measurements also imply a significant difference in the electronic structure between the two SC phases, highlighting the importance of electron carrier density in the high-$T_c$ phase. Our work devotes the effort to establish the commonality of pressure effect on the BiS$_2$-based superconductors, and calls on future detailed structural and theoretical investigations on this new SC phase under pressure.

%\section*{Acknowledgments}

We thank Tuson Park and A. M. Mounce for helpful technical support. Work at Los Alamos was performed under the auspices of the US Department of Energy, Division of Materials Science and Engineering. Y. Luo acknowledges a Director's Postdoctoral Fellowship supported through the Los Alamos LDRD program. Work at ZJU was supported by the National Basic Research Program of China (No. 2011CBA00103), and the Natural Science Foundation of China (No. 11190023).


\begin{thebibliography}{10}
\bibitem{Bednorz-La2CuO4_Sr}
J. G. Bednorz, and K. A. M\"{u}ller, Z. Phys. B \textbf{64}, 189 (1986).
\bibitem{Hosono-La1111_F}
Y. Kamihara, T. Watanabe, M. Hirano, and H. Hosono, J. Am. Chem. Soc. \textbf{130}, 3296 (2008).
\bibitem{Mizuguchi-Bi4O4S3}
Y. Mizuguchi, H. Fujihisa, Y. Gotoh, K. Suzuki, H. Usui, K. Kuroki, S. Demura, Y. Takano, H. Izawa, and O. Miura, Phys. Rev. B \textbf{86}, 220510 (2012).
\bibitem{Mizuguchi-La1121_F}
Y. Mizuguchi, S. Demura, K. Deguchi, Y. Takano, H. Fujihisa, Y. Gotoh, H. Izawa, and O. Miura, J. Phys. Soc. Jpn. \textbf{81}, 114725 (2012).
\bibitem{WenHH-Ce1121_F}
J. Xing, S. Li, X. Ding, H. Yang, and H. H. Wen, Phys. Rev. B \textbf{86}, 214518 (2012).
\bibitem{Jha-Pr1121_F}
R. Jha, A. Kumar, S. K. Singh, and V. P. S. Awana, J. Supercond. Nov. Magn. \textbf{26}, 499 (2013).
\bibitem{Yazici_Ln1121_F}
D. Yazici, K. Huang, B. D. White, A. H. Chang, A. J. Friedman, and M. B. Maple, Philos. Mag. \textbf{93}, 673 (2012).
\bibitem{Demura-Nd1121_F}
S. Demura, Y. Mizuguchi, K. Deguchi, H. Okazaki, H. Hara, T. Watanabe, S. J. Denholme, M. Fujioka, T. Ozaki, H. Fujihisa, Y. Gotoh, O. Miura, T. Yamaguchi, H. Takeya, and Y. Takano, J. Phys. Soc. Jpn. \textbf{82}, 033708 (2013).
\bibitem{Maziopa-LaSe1121_F}
A. Krzton-Maziopa, Z. Guguchia, E. Pomjakushina, V. Pomjakushin, R. Khasanov, H. Luetkens, P. K. Biswas, A. Amato, H. Keller, and K. Conder, J. Phys.: Condens. Matter \textbf{26}, 215702 (2014).
\bibitem{WenHH-LaSe1121pre}
J. Liu, S. Li, Y. Li, X. Zhu, and H. H. Wen, Phys. Rev. B \textbf{90}, 094507 (2014).
\bibitem{LiYK-Sr1121_La}
X. Lin, X. Ni, B. Chen, X. Xu, X. Yang, J. Dai, Y. Li, X. Yang, Y. Luo, Q. Tao, G. Cao, and Z. Xu, Phys. Rev. B \textbf{87}, 020504(R) (2013).
\bibitem{Jha-Sr1121_Ln}
R. Jha, B. Tiwari, and V. P. S. Awana, arXiv: 1407. 3105 (2014).
\bibitem{Zhai-Eu1121}
H. F. Zhai, Z. T. Tang, H. Jiang, K. Xu, K. Zhang, P. Zhang, J. K. Bao, Y. L. Sun, W. H. Jiao, I. Nowik, I. Felner, Y. K. Li, X. F. Xu, Q. Tao, C. M. Feng, Z. A. Xu, and G. H. Cao, Phys. Rev. B \textbf{90}, 064518 (2014).
\bibitem{Maple-LaCe1121pre}
C. T. Wolowiec, D. Yazici, B. D. White, K. Huang, and M. B. Maple, Phys. Rev. B \textbf{88}, 064503 (2013).
\bibitem{Tomita-La1121pre}
T. Tomita, M. Ebata, H. Soeda, H. Takahashi, H. Fujihisa, Y. Gotoh, Y. Mizuguchi, H. Izawa, O. Miura, S. Demura, K. Deguchi, and Y. Takano, J. Phys. Soc. Jpn. \textbf{83}, 063704 (2014).
\bibitem{Zhai-Eu3244}
H. F. Zhai, P. Zhang, S. Q. Wu, C. Y. He, Z. Tu. Tang, H. Jiang, Y. L. Sun, J. K. Bao, I. Nowik, I. Felner, Y. W. Zeng, Y. K. Li, X. F. Xu, Q. Tao, Z. A. Xu, and G. H. Cao, J. Am. Chem. Soc. \textbf{136}, 15386 (2014).
\bibitem{Maple-Ln1121pre}
C. T. Wolowiec, B. D. White, I. Jeon, D. Yazici, K. Huang, and M. B. Maple J. Phys.: Condens. Matter \textbf{25}, 422201 (2013).
\bibitem{Mizuguchi-La1121Hc2}
Y. Mizuguchi, A. Miyake, K. Akiba, M. Tokunaga, J. Kajitani, and O. Miura, Phys. Rev. B \textbf{89}, 174515 (2014).
\bibitem{Nagao-Ln1121SX}
M. Nagao, A. Miura, S. Demura, K. Deguchi, S. Watauchi, T. Takei, Y. Takano, N. Kumada, and I. Tanaka, Solid State Commun. \textbf{178}, 33 (2014).
\bibitem{Usui-band}
H. Usui, K. Suzuki, and K. Kuroki, Phys. Rev. B \textbf{86}, 220501(R) (2012).
%\bibitem{Kotegawa-BiS2pre}
%H. Kotegawa, Y. Tomita, H. Tou, H. Izawa, Y. Mizuguchi, O. Miura, S. Demura, K. Deguchi, and Y. Takano, J. Phys. Soc. Jpn. \textbf{81}, 103702 (2012).

\end{thebibliography}
\end{document}